\RequirePackage{fix-cm}
\documentclass[smallextended]{svjour3}       
\smartqed  
\usepackage{graphicx}
\usepackage{natbib}
\newcommand{\OurProject}{I2C}

\journalname{Experimental Astronomy}

\begin{document}

\title{Optical long baseline intensity interferometry\,:
prospects for stellar physics}



\titlerunning{Long baseline intensity interferometry} 

\author{Jean-Pierre Rivet
\and Farrokh Vakili
\and Olivier Lai 
\and David Vernet
\and Mathilde Fouch{\'e}
\and William Guerin
\and Guillaume Labeyrie
\and Robin Kaiser
}
\authorrunning{Rivet \emph{et al.}} 

\institute{Jean-Pierre Rivet
\and Farrokh Vakili
\and Olivier Lai 
\and David Vernet \at
  $^{1}$Universit{\'e} C{\^o}te d'Azur, OCA, CNRS, Lagrange, France \\
  \email{Jean-Pierre.Rivet@oca.eu}           
\and Mathilde Fouch{\'e}
\and William Guerin
\and Guillaume Labeyrie
\and Robin Kaiser \at
  $^{2}$Universit{\'e} C{\^o}te d'Azur, CNRS, Institut de Physique de Nice, France \\
}

\date{Submitted to Experimental Astronomy, January 31st, 2018}

\maketitle

\begin{abstract}
More than sixty years after the first intensity correlation experiments
by Hanbury~Brown and Twiss, there is renewed interest for intensity
interferometry techniques for high angular resolution studies of
celestial sources. We report on a successful attempt to measure the
bunching peak in the intensity correlation function for bright stellar
sources with $1$~meter telescopes (\OurProject\ project). We propose further
improvements of our preliminary experiments of spatial interferometry 
between two $1$~m telescopes, and discuss the possibility
to export our method to existing large arrays of telescopes.

\end{abstract}

\section{Introduction: historical background and revival}
\label{sec:Intro}

Like classical amplitude interferometry, intensity interferometry is 
a high-resolution observing technique used in astronomy to extend 
the angular resolution far beyond the classical resolution limit of 
single aperture instruments (optical or radio telescopes).

The principle underlying intensity interferometry is to monitor the
space and/or time correlations of the electromagnetic wave intensity fluctuations in a given, limited bandwidth. 

This idea was first proposed by Robert Hanbury~Brown and theorized by 
Richard Twiss in the early 1950', in order to estimate the sizes of
radio-sources beyond the resolution capabilities of classical amplitude
radio-interferometry. This method was first tested on-sky by~\citet{HBJDG:1952} to determine the
size of ``Cygnus~A'' and ``Cassiopea~A'' radio-sources.
Independently, similar diameter values were obtained for these objects
by classical amplitude interferometry by~\citet{Mills:1952} and~\citet{Smith:1952}.

In order to extend their concept to the optical domain, Hanbury~Brown and
Twiss (HBT hereafter) first performed successful laboratory experiments 
on an artificial light source (Mercury arc lamp) with two photomultiplier
tubes~\citep{HBT:1956a}. They demonstrated further the efficiency of their
method in real, less than ideal atmospheric conditions,
by measuring the diameter of the star Sirius with a prototype
intensity interferometer involving two $1.56$~m searchlight projectors
used as photons collectors~\citep{HBT:1956b}.

This first preliminary experiment paved the way for a more ambitious
experiment\,: the Narrabri stellar intensity interferometer (NSII).
The experimental setup involved two $6.5$~m reflectors on a 
circular track with diameter $188$~m. This configuration allowed
adjusting the interferometric baselines up to $188$~m, whilst
maintaining the baseline projection constant~\citep{HBT:1967a,HBT:1967b,HBT:1968}.
From 1964 to 1972, the NSII has been used to perform ambitious experiments
and led to the first consistent catalog of diameter measurements for $32$~stars from spectral type O5 to F8~\citep{HBrown:1974a}.

Classical amplitude interferometry requires to control the 
time-dependent sidereal optical path difference (OPD) between the arms of the
interferometer with accuracies better than the wavelength of the 
radiation under consideration, on sub-millisecond time scales. This 
constraint is not considered as a challenge anymore, at least for 
long wavelengths (radio-interferometers) or in the optical domain for
baselines of at most a few hundred meters, like
for VLTI, CHARA and NPOI astronomical interferometers 
\citep{Gomes:2017,Mourard:2015,Garcia:2016}. In the latter case, high
precision optical delay lines are required to transport coherently the 
light signal from individual telescopes to the focal plane of the
interferometer. For amplitude interferometry
with very long baselines at short wavelengths (optical domain), the
OPD compensation is still a challenging issue. 

On the contrary, intensity interferometry only requires OPD control with accuracies given by the bandwidth of the bandpass filter (equivalently, 
by the time resolution of the acquisition chain),
rather than by the wavelength of the radiation itself. For example,
an intensity acquisition chain with $100$~ps time-resolution requires 
an OPD control with millimetric accuracy only. Moreover, the link between
individual telescopes only involves signal cables and no optical delay lines.
In addition, the role of the telescopes is only to collect photons onto
the sensitive surfaces of the sensors. 
This requires correct guiding and if possible
fast tip-tilt correction, but the optical quality of the telescopes
can be less-than-ideal. Besides, as emphasized by~\citet{Tan:2016},
operation is still possible
in real conditions, that is, through atmospheric turbulence and even 
with some sky background contamination (full Moon).

Intensity interferometry is thus much easier to implement,
especially with long baselines and is less sensitive to optical
defects and atmospheric turbulence. However, it suffers from a lack of
sensitivity~\citep{HBlivre:1974},
requiring large collecting areas. For this reason, intensity interferometry
has not been developed further since the early 1970', until the last decade
where a significant jump was performed by the technology of light
detectors and signal digital correlators. Single photon avalanche diodes
(SPAD) are now commercially available in the visible domain, with
unprecedented sensitivity, short dead time and high time resolution. 
Moreover, time to digital converters (TDC) which can process the electric
pulses delivered by SPADs with count rates as high as several~Mcps (millions
of counts per second) and time resolution of a few picoseconds are now
also available. A review of commonly used SPADs and TDCs can be found
in table~1 from~\citet{Pilyavsky:2017}. These recent technological advances
shed a new light on intensity interferometry for astronomy. 
For this reason, and despite its lack of sensitivity, intensity
interferometry is now considered as a possible alternative to standard
amplitude interferometry for very long baselines (kilometric and more)
in the visible and near infrared domain. 

This solution has been proposed as an alternative observing mode for
the future Cherenkov Telescope
Array~\citep{LeBohec:2006,Nunez:2012,Dravins:2013}.
With several tens of $4$~m to $23$~m telescopes spread over
kilometric-sized areas, this facility will offer an unprecedented 
photon collecting area with sub-milliarcsecond angular resolutions
and dense $(u,v)$ plane coverage. However, these telescopes are not 
designed to be diffraction-limited and have a very large aperture ratios.
This is likely to introduce technical difficulties to couple the 
telescopes with the photon detection devices.

Another approach to revive intensity interferometry methods is to 
implement it on existing diffraction-limited optical telescopes.
This is the strategy followed for example by the ``Asiago
group''~\citep{Capraro:2010,Zampieri:2016}.
They have implemented intensity interferometry on the Copernicus and
Galileo telescopes in Asiago ($1.82$~m and $1.22$~m respectively),
coupled with the cutting edge high velocity and high efficiency 
photon-counting photometers ``Aqueye+'' and ``Iqueye''~\citep{Naletto:2009}.

In 2017, Pilyavsky \emph{et al.} conducted
extensive numerical simulations of intensity interferometry on seven
$0.9$~m to $4.0$~m telescopes at Kitt~Peak observatory~\citep{Pilyavsky:2017}.
Their simulations suggest that valuable science throughput on
stellar diameters of hot stars can be achieved with commercially
available components and without excessive technical complexity.

This renewal of interest around intensity interferometry has 
led our research team to conduct new experiments with state-of-the-art
avalanche photodiodes and time-to-digital converters. In 2016, we
had performed laboratory experiments with artificial light 
sources~\citep{Dussaux:2016}, to assess the capability of our setup
to detect the photon bunching phenomenon. In this article, we describe 
further experiments we have conducted in 2017 on bright stars with one,
then with two $1$~m telescopes. We also discuss further more ambitious
experiments that we plan to perform in a near future. We finally present
possible implementations of these techniques on existing arrays
of telescopes.

\section{From amplitude to intensity interferometry on the French Riviera}
\label{sec:AmplIntInterfero}

Following the pioneer optical long baseline interferometers on
the French Riviera~\citep{Labeyrie:1975,Labeyrie:1986} and given our 
currently operating focal instruments (VEGA-CHARA at 
Mount~Wilson~\citep{Mourard:2015}
and AMBER at Paranal ESO-VLTI~\citep{Petrov:2007}
or MATISSE~\citep{Lopez:2014}, recently commisioned), we have initiated 
pilot experiments in 2015, aimed at attempting intensity interferometry 
in the visible and the near infrared using the two $1$~meter optical
telescopes of
the C2PU facility~(https://www.oca.eu/index.php/fr/accueil-c2pu) 
in view of implementing this technique on
future arrays of optical telescopes. The C2PU facility is located at
the Calern observing site of the \emph{Observatoire de la C{\^o}te d'Azur}
(hereafter OCA).
We call this project ``\OurProject'' for Intensity Interferometry at Calern.
\OurProject\ follows a stepwise approach
to validate the focal instrument concept and components, including analysis
and interpretation methods, and to reliably estimate the achievable
sensitivity.

Like other groups, we have considered application to Cerenkov
arrays which can be used for intensity interferometry during bright time,
especially CTA for its photon-collecting surface which is almost equivalent
to a $100$~m dish with a maximum baseline of $2500$~m,
once the array is completed.
However the angular spread of a Cerenkov telescope focal image is large,
so that the light can only poorly be coupled to a single multimode fiber.
Also, the sky background photon noise, integrated over the entire Cerenkov
telescope angular spread, will affect the limiting magnitude. It is possible
to introduce  optical components to correct for the Cerenkov telescope primary
mirror aberrations, but this dramatically complicates the design and
complexity of these arrays primarily dedicated to high energy 
astrophysics. Another possible approach would be to use large surface detectors such as photomultipliers~\citep{Matthews:2017}, at the expense of a lower electronic bandpass, therefore lower intrinsic setup visibility.

Our technical choice is based on optical fibers connected to SPADs 
rather than photomultipliers. So, there are advantages to using 
existing optical telescopes to implement intensity
interferometry techniques. Optical telescopes, whatever their size, 
generally provide seeing limited focal images, so that stellar photons
can be transported by optical fibers to photon-counting detectors 
in order to estimate the intensity correlation function
$g^{2}(\tau,\rho)$ defined by\,:
\begin{equation}
  g^{(2)}(\tau,\rho) = \frac{<I(t,r)I(t+\tau,r+\rho)>}{<I(t,r)>^{2}}.
\end{equation}
With seeing-limited telescopes, at least $50\%$ coupling efficiencies
are achievable with multimode optical fibers, which are cheap and 
commercially available.
However, the very narrow bandpass required to preserve coherence of the
photons implies that a large fraction of the light is rejected
and not used in the correlation measurements. Therefore as already noted
by~\citet{HBlivre:1974}, the most promising way to significantly improve
the instrumental sensitivity of intensity interferometry is to observe
spectrally dispersed light from the source with multiple photodetectors
simultaneously at different wavelengths \emph{i.e.}, multiple spectral
channels. Besides accessing valuable spectral lines to probe stellar
surfaces and circumstellar structures, triggered by magnetic fields, 
mass loss or colliding winds, there is also gain on the intrinsic
instrumental contrast, and hence the signal-to-noise ratio (SNR), 
which improves as the square root of the spectral channels 
simultaneously correlated~\citep{Trippe:2014}. 

To keep the volume and design of a high dispersion spectrograph reasonable,
it is much preferable to use single mode (SM) fibers, which allow the use of
fibered Fabry-P{\'e}rot etalons, fibered Bragg gratings or integrated (on-chip)
spectrograph technology, to work on a large number of simultaneous narrow
band spectral channels. To couple the light into SM fibers, the telescopes
must be equipped with adaptive optics; however, the optical throughput
and the decreasing Strehl ratio at short wavelengths can negatively affect
the sensitivity, and a trade study is required to compare different SM
coupling technologies, such as photonics lanterns. These can be used to
split the light into a number of SM fibers, which can then each be 
dispersed spectrally. The output of all the SM fibers can then be 
co-added before or after detection of each spectral channel.

\subsection{Results from temporal bunching at C2PU}
\label{ssec:TemporalBunching}

On the nights of February 20, 21, and 22, 2017, we have
measured~\citep{Guerin:2017} what we believe to be the first intensity
correlation measured with the light of a star other than 
the sun~\citep{Tan:2014} since HBT
historical experiments. The temporal photon bunching $g^{(2)}(\tau, r = 0)$,
obtained in the photon counting regime, was measured for $3$ bright
stars, $\alpha$~Boo, $\alpha$~CMi, and $\beta$~Gem. The light was collected
at the focal plane of the West telescope of the C2PU facility 
(IAU observatory code\,: $010$,
latitude\,: $43^\circ\,\,45'\,\,13''$~N, longitude\,: 
$06^\circ\,\,55'\,\,22''$~E, altitude\,: $1270$~m). The instrument is a 
$F/12.5$ pure Cassegrain combination, but to improve the fiber coupling
efficiency, we have reduced the focal ratio to $F/5.6$ by inserting a
combination of commercially available focal reducers. In the resulting
focal plane, we have placed our compact optical setup. It consists of
a dichroic beam splitter, a CCD guiding camera, a combination of
filters and a multi-mode fiber (see Fig.~1 in~\citet{Guerin:2017}). The light 
reflected by the dichroic ($\lambda<650$~nm) is sent to the guiding camera.
The transmitted light goes through a linear polarizer, then through 
two cascaded narrow band dielectric filters centered on $\lambda_0=780$~nm.
The resulting bandwidth is $1$~nm with a throughput estimated as $61\%$
for the central wavelength.

The light is then transported by a $20$~m-long multimode graded-index
fiber (MMF) with a core diameter of $100\,\mu$m,  which is connected to 
a $50$:$50$ fibered splitter such that two photodetectors can be used 
to compute the autocorrelation function without being limited by the dead
time of the detectors, which is much larger than the coherence times. The
two output ports of the splitter are connected to SPADs
via two other MMFs of $1$~m for the first detection channel and $2$~m for 
the second. In order to avoid any 
spurious correlation induced by electronic cross-talk, we have introduced
an electronic delay between the two channels by using a $10$~m 
shielded BNC cable for the second channel and $1$~m for the first channel. 
The total delay (optical and electronic) between the two channels is 
$t_{0}\simeq45$~ns and is subtracted in the data processing.
For each detected photon, the SPADs produce a $10$~ns pulse, whose rising 
edge is detected and processed by a TDC with a
time bin of $162$~ps. The TDC is operated in the ``Delay-Histogram'' mode,
which yields a good approximation of the intensity correlation function after proper normalization.

\begin{figure*}
 \centerline{\includegraphics[width=0.75\textwidth]{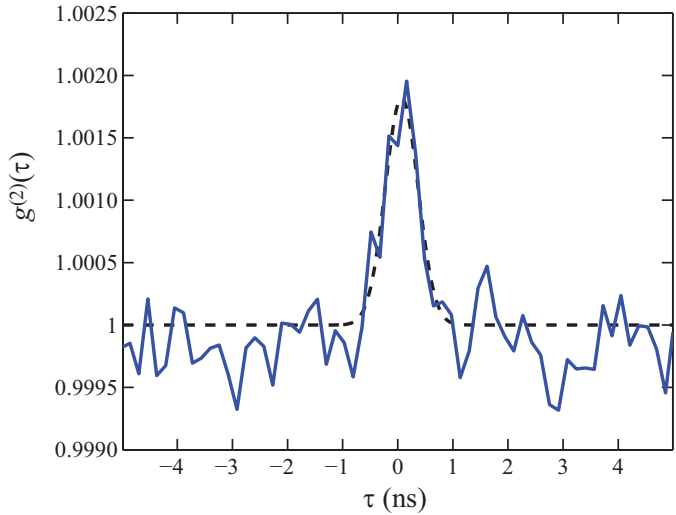}}
  \caption{Temporal intensity correlation function $g^{(2)}(\tau,r=0)$ 
  measured on $\alpha$~Boo (Arcturus) in February 2017. The Gaussian fit
  (dashed line) is used to estimate the contrast (reported in
  Table~\ref{tab:TemporalBunching}) and the FWHM of the bunching peak.
  The fit window is $[-10 , 10]$~ns.}
  \label{fig:g2_Arcturus}
\end{figure*}

For total exposure times of a few hours, we obtained contrast values around 
$2 \times 10^{-3}$ (see Table~\ref{tab:TemporalBunching}), in agreement 
with the theoretical expectation for chaotic sources, given the optical
and electronic bandwidths of our setup. Figure~\ref{fig:g2_Arcturus} displays
a typical correlation graph ($g^{(2)}(\tau)$ \emph{vs} $\tau$) obtained on
the star $\alpha$~Boo (Arcturus).

\begin{table*}
  \caption{Main data and results of temporal bunching on the three stars
  observed in February 2017 at C2PU.
  $T$ is the total integration time, in hours and minutes.
  The flux $\mathcal{F}$ is the number of detected photon counts per second
  and per detector, averaged over the total integration time.
  The contrast $\mathcal{C} = g^{(2)}(0)-1$ is the value of the correlation
  at zero delay given by the amplitude of the Gaussian fit,
  its uncertainty is the $1\sigma$ confidence interval of the fit. The noise
  is the \emph{rms} noise on the data evaluated in the wings of the
  correlation function, where  $g^{(2)}(\tau)\simeq 1$.}
  \label{tab:TemporalBunching}
  \begin{tabular}{llllllll}
    \hline
    Star & Sp. type & R mag & I mag & $T$ & 
    $\mathcal{F}$ ($\times 10^6$) &
    $\mathcal{C}$ $(\times 10^{-3})$ & noise $(\times 10^{-3})$  \\
    \hline
    $\alpha$ Boo & K0-III C & $-1.03$ & $-1.68 $ & 1:55' & 2.29 & 
    $1.81 \pm 0.19$ & 0.32\\
    $\alpha$ CMi & F5IV-V & $-0.05$ & $-0.28$ & 4:35' & 1.44 & 
    $1.90 \pm 0.28$ & 0.46\\
    $\beta$ Gem & K0-III & $+0.39 $ & $-0.11$ & 6:50' & 0.85 & 
    $2.38 \pm 0.43$ & 0.78\\
    \hline
  \end{tabular}
\end{table*}

It is worth comparing the performances of our measurements with the
Narrabri intensity interferometry results published by~\citet{HBT:1967b}. 
For this sake, we can use the star 
$\alpha$~CMi, which has been observed by both instruments. 
For the Narrabri instrument, the contrast of the correlation measured 
with the shortest baseline is reported with a SNR of $11.5$ for 
a $13.9$~hours observing time. We obtained a SNR of $6.8$ with only 
$4.6$~hours observing time and with a collecting area approximately 
$85$~times smaller (the Narrabri interferometer used two $6.5$~m 
collectors, whilst we used a single $1$~m telescope with $9.7\%$
central obscuration). The improvement in performance is due to the
larger electronic bandwidth, the better quantum efficiency of the
detectors, the fact that there are no correlation losses
due to the separation of the telescopes (the minimal baseline
of the Narrabri interferometer was $9.5$~m) and our working 
wavelength of $780$~nm, which is more adapted to
the spectral type (F5IV) of $\alpha$~CMi than Narrabri's
wavelength ($443$~nm).
Moreover, the performances of the system used by~\citet{Guerin:2017} can
be improved in two ways. First, a faster TDC and/or the choice of the 
``photon time tagging'' mode, associated to an appropriate data
post-processing system, allow for improved measurements of the correlation
function. Second, we can increase the efficiency of our telescope-to-fiber
coupling strategy by introducing a fast tip-tilt correction stage or,
preferably, a low-order adaptive optics. 

\subsection{Preliminary results from spatial bunching}
\label{ssec:g2deR}

The transition from the temporal bunching experiments of February
2017~\citep{Guerin:2017} to spatial bunching observations has been quite
fast and the first successful observations were performed in October~2017.
The presence in the same facility (C2PU) of two independent but nearly
identical $1$~m telescopes, separated by $15$~m has made this evolution
relatively easy.
For this, the fibered splitter used to feed both SPADs from the flux of a
single telescope was removed. Instead, two optical fibers coming from 
both telescopes were connected to the two SPADSs. So, our telescope-to-fiber
coupling system has been duplicated (same guiding camera, same set of filters)
and both replicas were set in the focal planes of the two telescopes. 

\begin{figure*} \centerline{\includegraphics[width=0.75\textwidth]{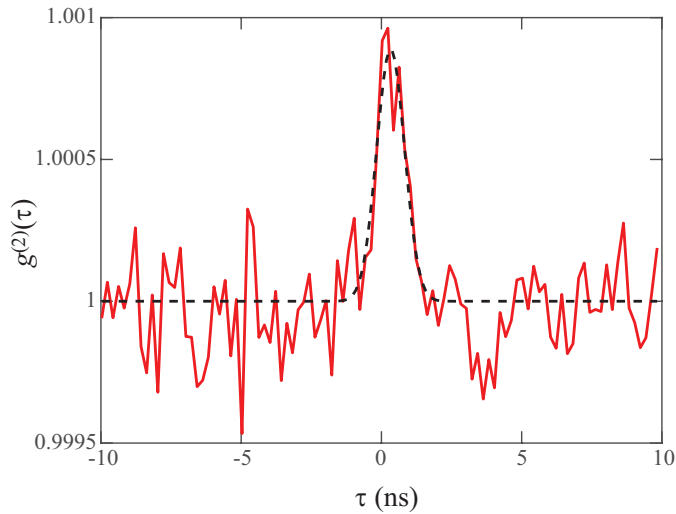}}
  \caption{Spatial intensity correlation function $g^{(2)}(\tau,r)$
  measured for $\alpha$~Lyr in October 2017. The Gaussian fit 
  (dark solid line) is used to extract the contrast and
  the FWHM of the bunching peak. The fit window is $[-10, 10]$~ns.}
  \label{fig:Vega_with_fit}
\end{figure*}

Three bright stars ($\alpha$~Lyr,  $\beta$~Ori and  $\alpha$~Aur) were
observed during $4$~nights, from $10^\textrm{th}$ to $13^\textrm{th}$ 
October 2017. The correlation peak emerged easily, provided a suitable delay
be taken into account (connection wire lengths difference and time-dependent
astronomical optical path difference).
Fig.~\ref{fig:Vega_with_fit} depicts the correlation peak detected 
on $\alpha$~Lyr for a total of $11.1$~hours integration time spread over
$3$~nights. The on-sky projected baseline ranged from $9.6$~m to $14.3$~m,
with a time-averaged value of $11.9$~m. The contrast was measured
as $(0.89\pm.09)\times10^{-3}$ ($1$-$\sigma$ uncertainty, \emph{i.e.}
SNR$=10$). Taking into account the temporal resolution of the detection chain,
this value is quite consistant with the known value of the uniform
disk angular diameter of $\alpha$~Lyr ($3.14$~mas), taking into account
a $11.9$~m projected baseline. 
A more detailed study on other bright stars
is under progress and will be presented in a forthcoming detailed paper.

\subsection{The \OurProject\ project\,: future developments}
\label{ssec:NextSteps}

Following the straightforward success of intensity interferometry
techniques on the C2PU facility at OCA, we have initiated a short
term plan to overcome its limitations and to improve its robustness
for a further deployment on larger facilities in other observatories
(see Sect.~\ref{sec:Discuss}).
The immediate plan is to improve our prototype experiment sub-system
components to gain on the sensitivity (limiting coherent magnitude,
\emph{i.e.} source apparent brightness and effective visibility) 
whilst comparing its performances to operating optical direct interferometers
like CHARA and/or NPOI through stellar sources with accurate angular
diameters.

A first step is based on introducing polarization beam-splitters to observe
stars with extended atmospheres and/or wind envelopes simultaneously in two
polarizations, parallel and perpendicular to the interferometric baseline. 
HBT's early observations at Narrabri on 
Rigel~\citep{HBrown:1974b} have proven that substantial gains in accuracy on
the visibility are required to measure any deviation from the uniform disk
model, according to realistic stellar atmosphere models of early type
stars~\citep{Cassinelli:1975}. 
Since then, direct interferometers have been unsuccessful to achieve such
measures~\citep{Vakili:1981} due to instrumental polarization effects
(\emph{e.g.} time-varying oblique reflections and cross-talk) introduced
by the optical complexity of amplitude interferometers. Intensity
interferometry is immune by construction to such effects since one can
directly separate two polarizations and correlate them independently without
any loss in SNR. Moreover, a gain of $\sqrt{2}$ is expected for unpolarized
sources, by co-adding the correlation functions of both polarization states.

\begin{figure*}
\centerline{\includegraphics[width=0.65\textwidth]{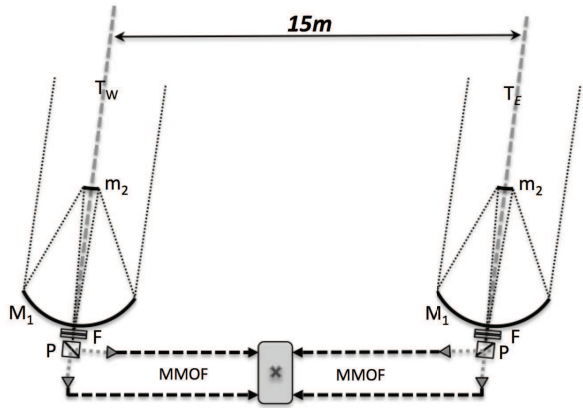}}
    \caption{Optical configuration of the \OurProject\ interferometer at OCA 
    for stellar polarization and/or differential angular diameter measurements
    in the continuum versus emission lines such as H$\alpha$. $T_W$ and 
    $T_E$ are respectively the West and East $1$~m optical telescopes of
    the C2PU facility. $M_1$ and $m_2$ are the primary and secondary mirrors
    (both telescopes have nearly identical optical combinations).
    Seeing limited images at the Cassegrain secondary foci are split in
    their two linearly polarized components and injected into two separate
    multimode optical fibers (MMOF), which transport the corresponding lights
    to SPADs. All four SPADs are connected to four input ports of the
    correlator $X$. Here, each polarization from both telescope is correlated
    to its counterpart from the other telescope so that $g^{(2)}(\tau,r)$
    estimates are obtained simultaneously in polarized light parallel and
    perpendicular to the baseline. For C2PU the maximum East-West baseline
    reaches $15$~m, corresponding to an angular resolution of $9.6$~mas at
    $700$~nm. Note that $F$ are $1$~nm bandwidth filters either centered on
    H$\alpha$ line or at the $780$~nm continuum to achieve differential
    visibility measurements.}
\label{fig:c2puii}
\end{figure*}

Fig.~\ref{fig:c2puii} depicts the generic configuration of the \OurProject\
interferometer project. This layout also allows to observe differential
effects between two spectral channels (one in the continuum and one
on some emission lines such as H$\alpha$). This is interesting for
hot stars, for instance Be or LBV like $\gamma$~Cas or P~Cyg, which
have been formerly observed in amplitude interferometry 
at OCA~\citep{Vakili:1997}.

A second step consists in increasing the angular resolution by
performing intensity correlation experiments involving one or two telescopes
from  C2PU and one telescope of the ``MeO'' laser-ranging
facility~\citep{Samain:2015}, located at $150$~m of
the C2PU building. Synchronization between these distant telescopes can be
achieved via cable or fiber connection with adequate control or measurement
of transmission delays. 

Most modern TDCs can be used in two modes\,: cross-correlation computation
and photon time-tagging mode. Both modes can be used for astronomical
intensity interferometry. If the electric signals of two (or more) SPADs 
linked to two (or more) telescopes, are fed into two (or more)
channels of the same TDC via coaxial cables, then the 
cross-correlation computation mode will deliver directly the photon
arrival time correlation function with a time delay which needs to 
be computed and specified \emph{a priori}. This operating mode requires
a physical connection between telescopes through cables.
On the contrary, if each
telescope has its own SPAD, its own TDC (set in time-tagging mode), which
delivers its time-tagging strings to its own computer, then no physical link
is required between the telescopes. Correlations can be computed numerically
afterwards. Of course, all TDCs need to be accurately synchronized to a common
time reference (\emph{e.g.} UTC).
This configuration is more expensive but more straightforward to extrapolate 
to very long baselines, since no physical connection is required. 
Another advantage is that the correlation
functions can be computed numerically \emph{a posteriori} from the sets of 
photon arrival times, recorded independently by different TDCs and stored
on independent mass storage devices. Contrary to correlations computed
``on the fly'' during the observation run, \emph{a posteriori} correlations
computation allows for more elaborate algorithms, for example with
time-varying and/or adjustable parameters.
This configuration with no physical link between telescopes could be tested
as a third step of our experiment. In our case, the necessary time reference
for the TDCs would be provided by the local time standard of the ``MeO''
instrument.
This time standard offers a temporal resolution and stability on the order of
a few picoseconds, which is significantly better than the coherence length of
our intensity correlation measurements (of the order of $3$~cm, because of the time-resolution of our SPADs). 

Finally, we are also considering $3$-telescope intensity correlation
experiments between the two C2PU $1$~m telescopes and the MeO laser-ranging
$1.5$~m telescope. The West-North-West direction of MeO with respect to 
C2PU offers much longer projected baselines than C2PU itself.
They range between a few tens of meters and $150$~m depending on the 
target's position. This leads to sub-mas resolutions for visible
wavelengths (\emph{e.g.} in the range of $600$-$800$~nm). 
Indeed, triple correlations between 
$3$~telescopes would provide the closure phase quantity~\citep{Malvimat:2014}
but at the expense of a much higher exposure time for a given SNR.
Several hundreds hours of integration would be necessary with the
aforementioned
OCA telescopes to attain the SNR value required to access any measure of
phase information from the triple correlation quantity. However the power 
of intensity interferometry, despite its intrinsic low SNR and sensitivity,
lies in the fact that the correlation signal can be integrated over many
nights since no fringe detection or tracking is necessary as in amplitude
interferometry.

\section{Prospects and discussion}
\label{sec:Discuss}

The revival of intensity interferometry half a century 
after the pioneering work by Hanbury~Brown and Twiss
suggests that these techniques are promised to open new opportunities for 
very high angular resolution observations in the future, specially in the
visible wavelengths, down to the violet if the site transparency permits.
Two approaches have
been proposed until now\,: the first one is based upon the use of
existing or foreseen Cerenkov telescope
arrays~\citep{Dravins:2008b,Matthews:2017},
and the second one is to use existing seeing limited optical
telescopes~\citep{Zampieri:2016,Pilyavsky:2017,Guerin:2017}. 
Simulations and laboratory
experiments confirm the imaging potential of HBT arrays whatever
Cerenkov or optical. We started and effectively demonstrated that photon
bunching, either temporal or spatial can be straightforwardly
achieved with two modest $1$~m telescopes at OCA using almost
``off-the-shelf'' quantum optics photonic components, at least if
instrumental effects are correctly controlled and calibrated.
As argued by \citet{Trippe:2014}, multichannel correlations, with
hundreds of simultaneous narrow spectral channels, is the most
promising way to progress over the intrinsic limitations of intensity
interferometry for a given limited number and photon collecting
surface. Much experience is to be gained however by
using existing $1$-$2$~m class telescopes, coupled later to $8$-$10$~m
class telescopes for on-sky observations in order to control the detailed
behavior of photonic subsystems and components of an intensity interferometer
on various astrophysical sources before speculating on the observation of
exotic or extragalactic sources with microarcsecond angular resolutions.
Considering this, we propose to consider intensity interferometry on a few
major observatories (see Table~\ref{tab:Sites}) and a first prototype
HBT array based on the four $1.8$~m ATs at Paranal/Chile (A.~M{\'e}rand, private
communication) that would be available for such experiments and observations
when the VLTI focal instruments MATISSE and GRAVITY occupy the Paranal
optical delay lines.

\begin{table*}
  \caption{Potential arrays of optical telescopes with the highest
  photon-collecting surfaces and kilometric baselines suitable for
  modern intensity interferometry. $N_{T}$ is the number of telescopes
  with diameters larger than $1.5$~m, $S$ is the total collecting surface
  of the array in $m^2$ (assuming $10\%$ to $15\%$ losses due to the central
  obstruction), $N_{b}$ is the number of possible array baselines, $B_{m}$
  is the array theoretical maximum baseline in meters and $\mathcal{\rho}$ 
  the potential angular resolution in mas at $700$~nm wavelength.
  Last column gives the potential photon collecting area and/or the longest
  baseline becoming available if extremely large telescopes are built
  at or close to the corresponding observing sites.
  }
  \label{tab:Sites}
  \begin{tabular}{lllllll}
    \hline
    Observatory & $N_{T}$ & $S$ & $N_{b}$ & $B_{m}$  &  
    $\mathcal{\rho}$ &     Comment \\
    \hline
    ESO Paranal & $8$ & $220$ & $28$ & $220$ & $0.7$ & $1200\,m^2$ 
    and $21$~km with ELT \\
    Mauna Kea & $8$ & $260$ & $28$ &  $2500$ & $0.06$ & 
    $1260\,m^2$ with TMT \\
    La Palma & $4$ & $100$ & $6$ & $1200$ & $0.12$ & 
    $1100\,m^2$ with TMT\\
    \hline
  \end{tabular}
\end{table*}

\begin{acknowledgements}
The \OurProject\ pilot experiment is supported by INPHYNI and
Lagrange laboratories, D{\"o}blin Federation and grants from
OCA and the Excellence Initiative UCA-JEDI from University
C{\^o}te d'Azur. We are grateful to A.~Dussaux for his valuable
contribution to this project. We also thank E.~Samain, C.~Courde
and J.~Chab{\'e} (GeoAzur lab., OCA) for fruitful discussions
about space and time metrology. Ph.~B{\'e}rio from Lagrange Laboratory
(OCA) is also kindly acknowledged.
\end{acknowledgements}

\bibliographystyle{spbasic}      


\begin{thebibliography}{37}
\providecommand{\natexlab}[1]{#1}
\providecommand{\url}[1]{{#1}}
\providecommand{\urlprefix}{URL }
\expandafter\ifx\csname urlstyle\endcsname\relax
  \providecommand{\doi}[1]{DOI~\discretionary{}{}{}#1}\else
  \providecommand{\doi}{DOI~\discretionary{}{}{}\begingroup
  \urlstyle{rm}\Url}\fi
\providecommand{\eprint}[2][]{\url{#2}}

\bibitem[{Capraro et~al(2010)Capraro, Barbieri, Naletto, Occhipinti, Verroi,
  Zoccarato, , and Gradari}]{Capraro:2010}
Capraro I, Barbieri C, Naletto G, Occhipinti T, Verroi E, Zoccarato P, ,
  Gradari S (2010) Quantum astronomy with {Iqueye}. Proc SPIE 7702:77,020M

\bibitem[{Cassinelli and Hoffman(1975)}]{Cassinelli:1975}
Cassinelli JP, Hoffman NM (1975) The effect of linearly polarized light from
  extended stellar atmospheres on interferometer response functions. MNRAS
  173:789--800

\bibitem[{Dravins and LeBohec(2008)}]{Dravins:2008b}
Dravins D, LeBohec S (2008) Toward a diffraction-limited square-kilometer
  optical telescope\,: digital revival of intensity interferometry. Proc SPIE
  6986:698,609

\bibitem[{Dravins et~al(2013)Dravins, LeBohec, Jensen, and
  {Nu{\~n}ez}}]{Dravins:2013}
Dravins D, LeBohec S, Jensen H, {Nu{\~n}ez} PD (2013) Optical intensity
  interferometry with the {Cherenkov Telescope Array}. Astropart Phys
  43:331--447

\bibitem[{Dussaux et~al(2016)Dussaux, {Passerat de Silans}, Guerin, Alibart,
  Tanzilli, Vakili, and Kaiser}]{Dussaux:2016}
Dussaux A, {Passerat de Silans} T, Guerin W, Alibart O, Tanzilli S, Vakili F,
  Kaiser R (2016) Temporal intensity correlation of light scattered by a hot
  atomic vapor. Phys Rev A 93:043,826

\bibitem[{Garcia et~al(2016)Garcia, Muterspaugh, {van~Belle}, Monnier, Stassun,
  Ghasempour, Clark, Zavala, Benson, Hutter, and { et al.}}]{Garcia:2016}
Garcia EV, Muterspaugh MW, {van~Belle} G, Monnier JD, Stassun KG, Ghasempour A,
  Clark JH, Zavala RT, Benson JA, Hutter DJ, { et al} (2016) {VISION}: {A
  Six-telescope Fiber-fed Visible Light Beam Combiner for the Navy Precision
  Optical Interferometer}. PASP 128:{055}{004}

\bibitem[{Gomes et~al(2017)Gomes, Garcia, and Thi{\'e}baut}]{Gomes:2017}
Gomes N, Garcia PJV, Thi{\'e}baut E (2017) Assessing the quality of restored
  images in optical long-baseline interferometry. MNRAS 465:3823--3839

\bibitem[{Guerin et~al(2017)Guerin, Dussaux, Fouch{\'e}, Labeyrie, Rivet,
  Vernet, Vakili, and Kaiser}]{Guerin:2017}
Guerin W, Dussaux A, Fouch{\'e} M, Labeyrie G, Rivet JP, Vernet D, Vakili F,
  Kaiser R (2017) Temporal intensity interferometry: photon bunching in three
  bright stars. MNRAS 472:4126--4132

\bibitem[{{Hanbury~Brown}(1968)}]{HBT:1968}
{Hanbury~Brown} R (1968) Stellar interferometer at {Narrabri} observatory.
  Nature 218:637--641

\bibitem[{{Hanbury~Brown}(1974)}]{HBlivre:1974}
{Hanbury~Brown} R (1974) The intensity interferometer\,: Its application to
  astronomy. Taylor and Francis, Ltd

\bibitem[{{Hanbury~Brown} and Twiss(1956{\natexlab{a}})}]{HBT:1956a}
{Hanbury~Brown} R, Twiss RQ (1956{\natexlab{a}}) Correlation between photons in
  two coherent beams of light. Nature 177:27--29

\bibitem[{{Hanbury~Brown} and Twiss(1956{\natexlab{b}})}]{HBT:1956b}
{Hanbury~Brown} R, Twiss RQ (1956{\natexlab{b}}) A test of a new type of
  stellar interferometer on {Sirius}. Nature 178:1046--1048

\bibitem[{{Hanbury~Brown} et~al(1952){Hanbury~Brown}, Jennison, and
  Gupta}]{HBJDG:1952}
{Hanbury~Brown} R, Jennison RC, Gupta MKD (1952) Apparent angular sizes of
  discrete radio sources: Observations at {Jodrell Bank, Manchester}. Nature
  170:1061--1063

\bibitem[{{Hanbury~Brown} et~al(1967{\natexlab{a}}){Hanbury~Brown}, Davis, and
  Allen}]{HBT:1967a}
{Hanbury~Brown} R, Davis J, Allen LR (1967{\natexlab{a}}) The stellar
  interferometer at {Narrabri} observatory -- {I}. MNRAS 137:375--392

\bibitem[{{Hanbury~Brown} et~al(1967{\natexlab{b}}){Hanbury~Brown}, Davis,
  Allen, and Rome}]{HBT:1967b}
{Hanbury~Brown} R, Davis J, Allen LR, Rome JM (1967{\natexlab{b}}) The stellar
  interferometer at {Narrabri} observatory -- {II}. MNRAS 137:393--417

\bibitem[{{Hanbury~Brown} et~al(1974{\natexlab{a}}){Hanbury~Brown}, Davis, and
  Allen}]{HBrown:1974a}
{Hanbury~Brown} R, Davis J, Allen LR (1974{\natexlab{a}}) The angular diameters
  of $32$~stars. MNRAS 167:121--136

\bibitem[{{Hanbury~Brown} et~al(1974{\natexlab{b}}){Hanbury~Brown}, Davis, and
  Allen}]{HBrown:1974b}
{Hanbury~Brown} R, Davis J, Allen LR (1974{\natexlab{b}}) An attempt to detect
  a corona around beta {Orionis} with an intensity interferometer using
  linearly polarized light. MNRAS 168:93--100

\bibitem[{Labeyrie(1975)}]{Labeyrie:1975}
Labeyrie A (1975) Interference fringes obtained on {Vega} with two optical
  telescopes. ApJ 196:L71--L75

\bibitem[{Labeyrie et~al(1986)Labeyrie, Schumacher, Dugu{\'e}, Thom, and
  Bourlon}]{Labeyrie:1986}
Labeyrie A, Schumacher G, Dugu{\'e} M, Thom C, Bourlon P (1986) Fringes
  obtained with the large `boules' interferometer at {CERGA}. A\&A 162:359--364

\bibitem[{LeBohec and Holder(2006)}]{LeBohec:2006}
LeBohec S, Holder J (2006) Optical intensity interferometry with atomospheric
  {Cherenkov} telescope array. ApJ 649:399--405

\bibitem[{Lopez et~al(2014)Lopez, Lagarde, Jaffe, Petrov, Sch{\"o}ller,
  Antonelli, Beckman, Berio, Bettonvil, Graser, and {et al.}}]{Lopez:2014}
Lopez B, Lagarde S, Jaffe W, Petrov R, Sch{\"o}ller M, Antonelli P, Beckman U,
  Berio P, Bettonvil F, Graser U, {et al} (2014) {MATISSE} status report and
  science forecast. Proc SPIE 9146:91,460Z

\bibitem[{Malvimat et~al(2014)Malvimat, Wucknitz, and Saha}]{Malvimat:2014}
Malvimat V, Wucknitz O, Saha P (2014) Intensity interferometry with more than
  two detectors\,? MNRAS 437:798--803

\bibitem[{Matthews et~al(2017)Matthews, Kieda, and LeBohec}]{Matthews:2017}
Matthews N, Kieda D, LeBohec S (2017) Development of a digital astronomical
  intensity interferometer\,: laboratory results with thermal light. J Mod Opt
  65:1336--1344

\bibitem[{Mills(1952)}]{Mills:1952}
Mills BY (1952) Apparent angular sizes of discrete radio sources: Observations
  at {Sydney}. Nature 170:1063--1064

\bibitem[{Mourard et~al(2015)Mourard, Monnier, Meilland, Gies, Millour,
  Benisty, Che, Grundstrom, Ligi, Schaefer, and { et al.}}]{Mourard:2015}
Mourard D, Monnier JD, Meilland A, Gies D, Millour F, Benisty M, Che X,
  Grundstrom ED, Ligi R, Schaefer G, { et al} (2015) Spectral and spatial
  imaging of the {Be+sdO} binary {$\phi$ Persei}. A\&A 577:A51

\bibitem[{Naletto et~al(2009)Naletto, Barbieri, Occhipinti, Capraro, {Di
  Paola}, Facchinetti, Verroi, Zoccarato, Anzolin, Belluso, and { et
  al.}}]{Naletto:2009}
Naletto G, Barbieri C, Occhipinti T, Capraro I, {Di Paola} A, Facchinetti C,
  Verroi E, Zoccarato P, Anzolin G, Belluso M, { et al} (2009) {Iqueye}, a
  single photon-counting photometer applied to the {ESO} new technology
  telescope. A\&A 508:531--539

\bibitem[{{Nu{\~n}ez} et~al(2012){Nu{\~n}ez}, Holmes, Kieda, and
  LeBohec}]{Nunez:2012}
{Nu{\~n}ez} PD, Holmes R, Kieda D, LeBohec S (2012) High angular resolution
  imaging with stellar intensity interferometry using air {Cherenkov} telescope
  array. MNRAS 419:172--183

\bibitem[{Petrov et~al(2007)Petrov, Malbet, Weigelt, Antonelli, Beckmann,
  Bresson, Chelli, Dugu{\'e}, Duvert, Gennari, Gl{\"u}ck, KernP., Lagarde,
  Le~Coarer, Lisi, Millour, Perraut, Puget, Rantakyr{\"o}, and { et
  al.}}]{Petrov:2007}
Petrov RG, Malbet F, Weigelt G, Antonelli P, Beckmann U, Bresson Y, Chelli A,
  Dugu{\'e} M, Duvert G, Gennari S, Gl{\"u}ck L, KernP, Lagarde S, Le~Coarer E,
  Lisi F, Millour F, Perraut K, Puget P, Rantakyr{\"o} F, { et al} (2007)
  {AMBER}, the near-infrared spectro-interferometric three-telescope {VLTI}
  instrument. A\&A 464:1--12

\bibitem[{Pilyavsky et~al(2017)Pilyavsky, Mauskopf, Smith, Schroeder, Sinclair,
  {van~Belle}, Hinkel, and Scowen}]{Pilyavsky:2017}
Pilyavsky G, Mauskopf P, Smith N, Schroeder E, Sinclair A, {van~Belle} GT,
  Hinkel N, Scowen P (2017) Single-photon intensity interferometry ({SPIIFy}):
  utilizing available telescopes. MNRAS 467:3048--3055

\bibitem[{Samain(2015)}]{Samain:2015}
Samain E (2015) Clock comparison based on laser ranging technologies. Int J Mod
  Phys D 24:1530,021

\bibitem[{Smith(1952)}]{Smith:1952}
Smith FG (1952) Apparent angular sizes of discrete radio sources: Observations
  at {Cambridge}. Nature 170:1065

\bibitem[{Tan et~al(2014)Tan, Yeo, Poh, Chan, and Kurtsiefer}]{Tan:2014}
Tan PK, Yeo GH, Poh HS, Chan AH, Kurtsiefer C (2014) Measuring temporal photon
  bunching in backbody radiation. ApJ 789:L10

\bibitem[{Tan et~al(2016)Tan, Chan, and Kurtsiefer}]{Tan:2016}
Tan PK, Chan AH, Kurtsiefer C (2016) Optical intensity interferometry through
  atmospheric turbulence. MNRAS 457:4291

\bibitem[{Trippe et~al(2014)Trippe, Kim, Lee, Choi, Oh, Lee, Yoon, Im, and
  Park}]{Trippe:2014}
Trippe S, Kim JY, Lee B, Choi C, Oh J, Lee T, Yoon SC, Im M, Park YS (2014)
  Optical multi-channel intensity interferometry -- or\,: How to resolve
  {O-stars} in the {Magellanic} clouds. Journal of the Korean Astronomical
  Society 47:235--253

\bibitem[{Vakili(1981)}]{Vakili:1981}
Vakili F (1981) Study of stellar polarization with the {CERGA} interferometer.
  A\&A 101:352--355

\bibitem[{Vakili et~al(1997)Vakili, Mourard, Bonneau, and Stee}]{Vakili:1997}
Vakili F, Mourard D, Bonneau D, Stee P (1997) Subtle structures in the wind of
  {P~Cygni}. A\&A 323:183--188

\bibitem[{Zampieri et~al(2016)Zampieri, Naletto, Barbieri, Barbieri, Verroi,
  Umbriaco, Favazza, Lessio, and Farisato}]{Zampieri:2016}
Zampieri L, Naletto G, Barbieri C, Barbieri M, Verroi E, Umbriaco G, Favazza P,
  Lessio L, Farisato G (2016) Intensity interferometry with {Aqueye+} and
  {Iqueye} in {Asiago}. Proc SPIE 9907:99,070N

\end{thebibliography}

%
%


\end{document}